\documentclass[pra,twocolumn,superscriptaddress]{revtex4} %preprint,
\usepackage{epsfig}

\usepackage{graphicx}% Include figure files

\usepackage{color,colordvi,graphicx,pstcol}
\newcommand{\myem}[1]{{\sc\large\color{red} #1}} %\color{red}
\newcommand{\ket}[1]{\ensuremath{\left| #1 \right\rangle}}

\newcommand{\tpaper}{this Letter} %this letter

\newcommand{\beq}{\begin{equation}}
\newcommand{\eeq}{\end{equation}}

\newcommand{\commentout}[1]{}

\begin{document}
\title{Diffusion induced decoherence of stored optical vortices}

\author{T. Wang}
\affiliation{Department of Physics, University of Connecticut,
Storrs, CT 06269, USA}
\author{L. Zhao}
\affiliation{Department of Physics, University of Connecticut,
Storrs, CT 06269, USA}
\author{L. Jiang}
\affiliation{Department of Physics, Harvard University, Cambridge,
Massachusetts 02138, USA}
\author{S. F. Yelin}
\affiliation{Department of Physics, University of Connecticut,
Storrs, CT 06269, USA} \affiliation{ITAMP, Harvard-Smithsonian
Center for Astrophysics, Cambridge, MA 02138, USA}

\date{\today}
\begin{abstract}
We study the coherence properties of optical vortices stored in
atomic ensembles. In the presence of thermal diffusion, the
topological nature of stored optical vortices is found not to
guarantee slow decoherence. Instead the stored vortex state has
decoherence surprisingly larger than the stored Gaussian mode.
Generally, the less phase gradient, the more robust for stored
coherence against diffusion. Furthermore, calculation of coherence
factor shows that the center of stored vortex becomes completely
incoherent once diffusion begins and, when reading laser is
applied, the optical intensity at the center of the vortex becomes
nonzero. Its implication for quantum information is discussed.
Comparison of classical diffusion and quantum diffusion is also
presented.
\end{abstract}
%\thanks{University of Connecticut}
 \maketitle

{\em Introduction} Photons can carry orbital angular momentum
(OAM), which can be created \cite{OpticalVortex0}, manipulated
\cite{Akamatsu03} and detected \cite{Leach02}. The OAM states are
associated with vortices of a helical phase structure
$e^{im\phi}$. Each vortex is a topological defect, characterized
by a winding number $m$, obtained from the $2\pi m$ phase twist
around the vortex. The light with OAM is exemplified by the most
commonly known Laguerre-Gaussian (LG) modes LG$_{p}^{m}$ with $p$
the number of nodes in radial direction~\cite{Molina01,LGintro}.
Entanglement through the OAM degree of freedom has been
demonstrated for photon pairs from parametric down conversion
\cite{Mair01}. Since the winding number can be any integers, the
OAM states have been proposed theoretically for multiple-alphabet
quantum cryptography with higher information density and a higher
margin of security \cite{Bechmann00,Groblacher06}. Other
proposals, such as quantum coin tossing and violation of Bell
inequalities, using the OAM states have been reviewed by Ref.
\cite{MolinaTerriza07}.

One of the key challenges for optically based quantum information
(QI) processing is the difficulty of storing optical fields. It
has been demonstrated~\cite{Dutton04,Kapale05,Andersen06} that a
superposition of the OAM\ states can be stored in a non-rotating
BEC in terms of vortex states of the condensate. Meanwhile, the
OAM states can also be stored in ``hot" atomic ensembles using
slow light
techniques~\cite{Lukin03,Flei05,StorageInVapor,ObservationStorage,VortexDiffusion}.
The information of photonic states, namely the amplitude and
phase, is continuously transformed into Raman coherence, i.e.,
spin-density waves, of the atomic ensemble, which can be later
retrieved. In practice, processes such as inhomogeneous magnetic
field~\cite{DuanMag02} and/or thermal diffusion can lead to the
decay of the Raman coherence. Then, to what extent can the optical
vortex states be stored coherently? Are they going to be more
robust than the Gaussian state? The answers to these questions
will determine how the OAM are used in QI. \commentout{ \myem{One
goal of \tpaper\ is to show what happens to these proposals in a
realistic situation (i.e., diffusion).}}

Generally speaking, the topological structure of vortices makes it
a good candidate for
QI~\cite{VortexQubit,TopoQubitJosephonJ,CNOTvortex} because it is
stable against continuous deformations which cannot cause it to
decay or to ``unwind".  %[Ann.Phys.303, 2 (2003)] scheme.
Actual studies of such robustness against various processes are of
course necessary. In particular, one needs to study the robustness
in the presence of diffusion~\cite{Yanhong} for the stored optical
vortex state, which is crucial for applications such as quantum
repeaters~\cite{QComm1,LiangRepeater07} and multiple beam
splitters for generating entangled single photons~\cite{Tun04}.
Towards answering this question, Pugatch, et
al.~\cite{VortexDiffusion} showed that in the presence of
diffusion, the dark center of a stored OAM mode is well preserved,
and the dark center of a Gaussian mode generated by blocking its
center disappears after a short time of diffusion. They attributed
the stable dark center of the stored OAM mode to the robustness of
the topological nature of vortex in the presence diffusion.

In this Letter, we provide a careful study for the robustness of
the stored vortex states (exemplified by $p=0$ unless otherwise
stated) in the presence of diffusion. We find that the stable dark
center of the vortex states is not directly associated with the
topological robustness. And the vortex states are actually more
vulnerable to diffusion than the Gaussian state. This is because:
(1) the diffusion is a global process during the storage and it
can destroy the topological order; (2) the readout process only
measures a specific output modes corresponding to Raman coherence
but not the full state;\commentout{in the memory;} (3) for vortex
states, the Raman coherence interferes destructively, while the
full quantum state evolves differently.

Our basic assumption is that there is no spin exchange or
interaction between atoms, the evolution of coherence obeys
diffusion equation $\dot{\rho}=D\nabla^2\rho$, where $\rho$ and
the $D$ are the density matrix and the diffusion coefficient,
respectively. We consider a three-level lambda system as generally
used for light storage~\cite{Lukin03,Flei05}: the weak probe laser
is applied between the ground state \ket1 and the excited state
\ket3, and the pump laser is addressing the transition between
\ket3 and another ground state \ket2.

\begin{figure}[t]
\centerline{\includegraphics[clip,width=.7\linewidth]{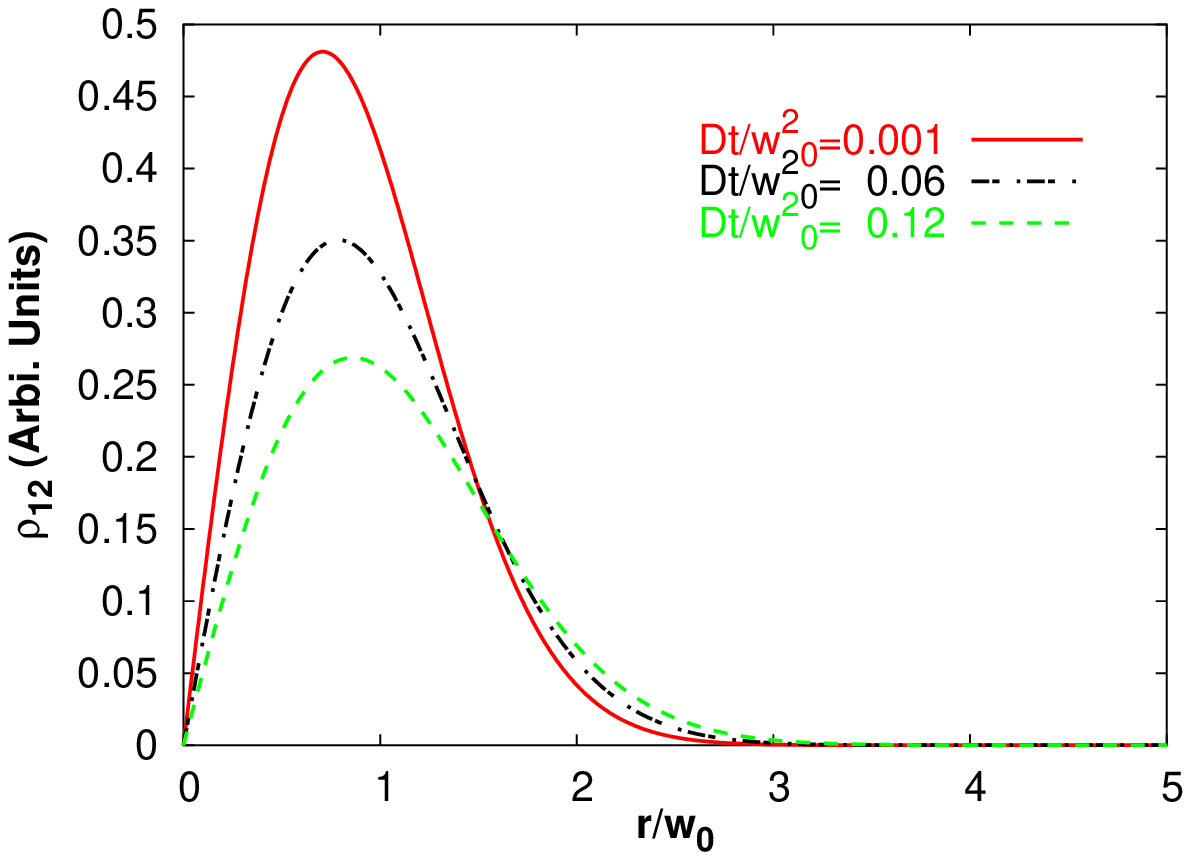}}
\centerline{\includegraphics[clip,width=.7\linewidth]{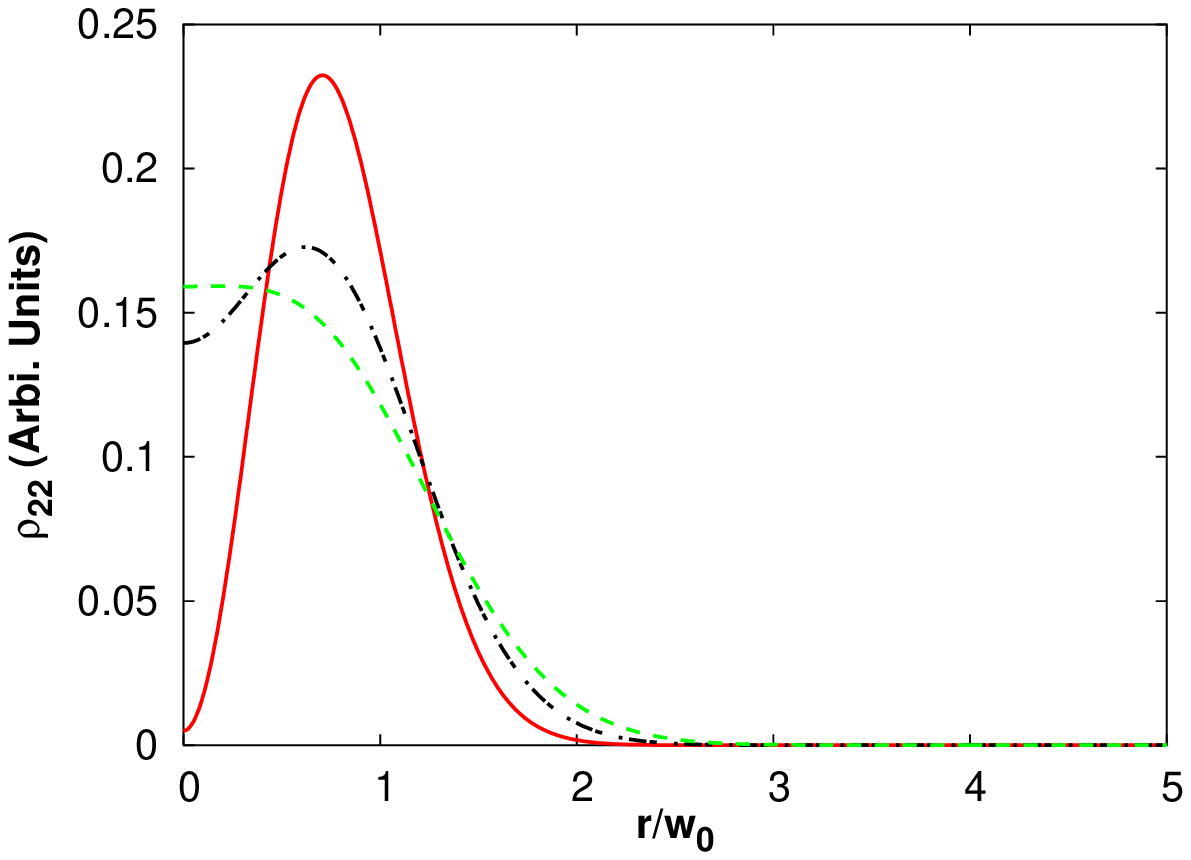}}
\centerline{\includegraphics[clip,width=.7\linewidth]{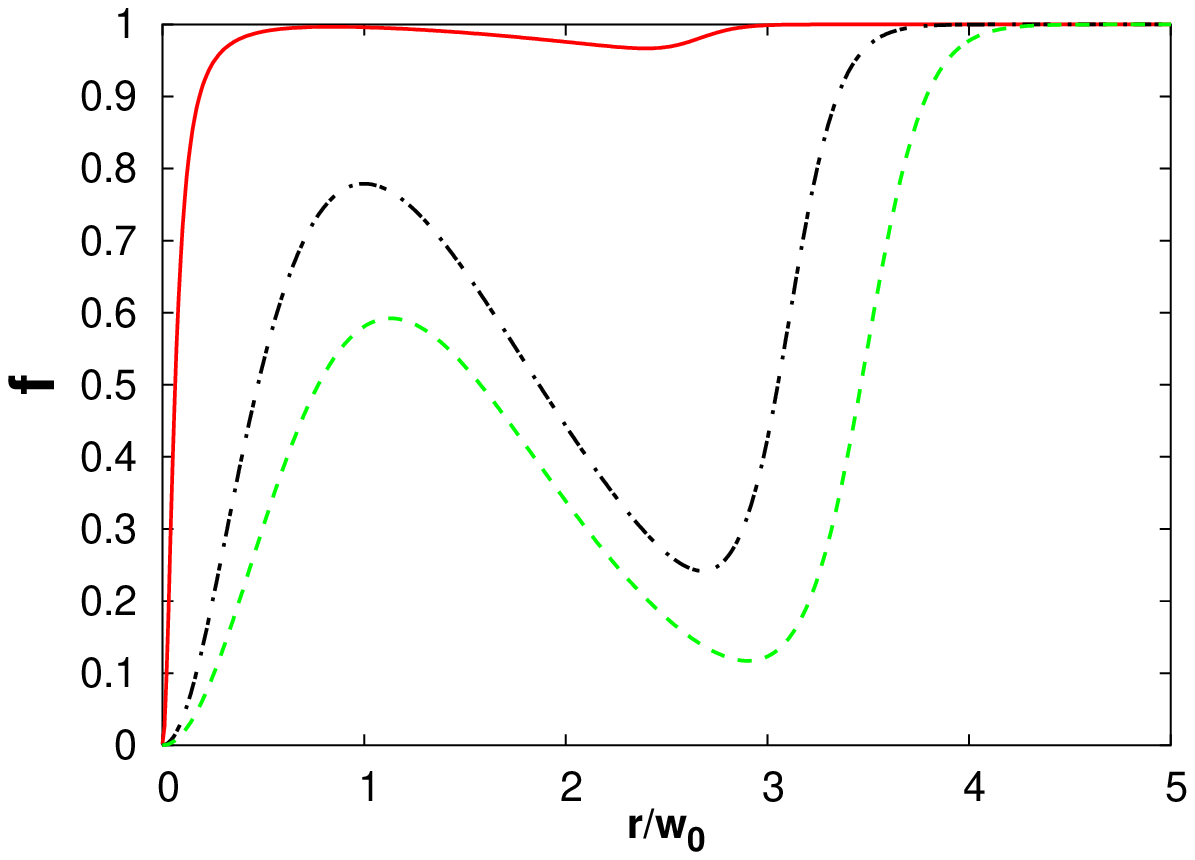}}
        \caption{\protect\label{vortex}  (Color online) Effects of
diffusion on a stored vortex with topological charge $m=1$.
Plotted are coherence $\rho_{12}$ (a), population $\rho_{22}$ (b)
and coherence factor $f$ (c) as functions of radius, for different
diffusion times. $w_0$, $D$ are the waist of the LG mode and the
diffusion coefficient, respectively.}
\end{figure}

\begin{figure}[t]
\centerline{\includegraphics[clip,width=.7\linewidth]{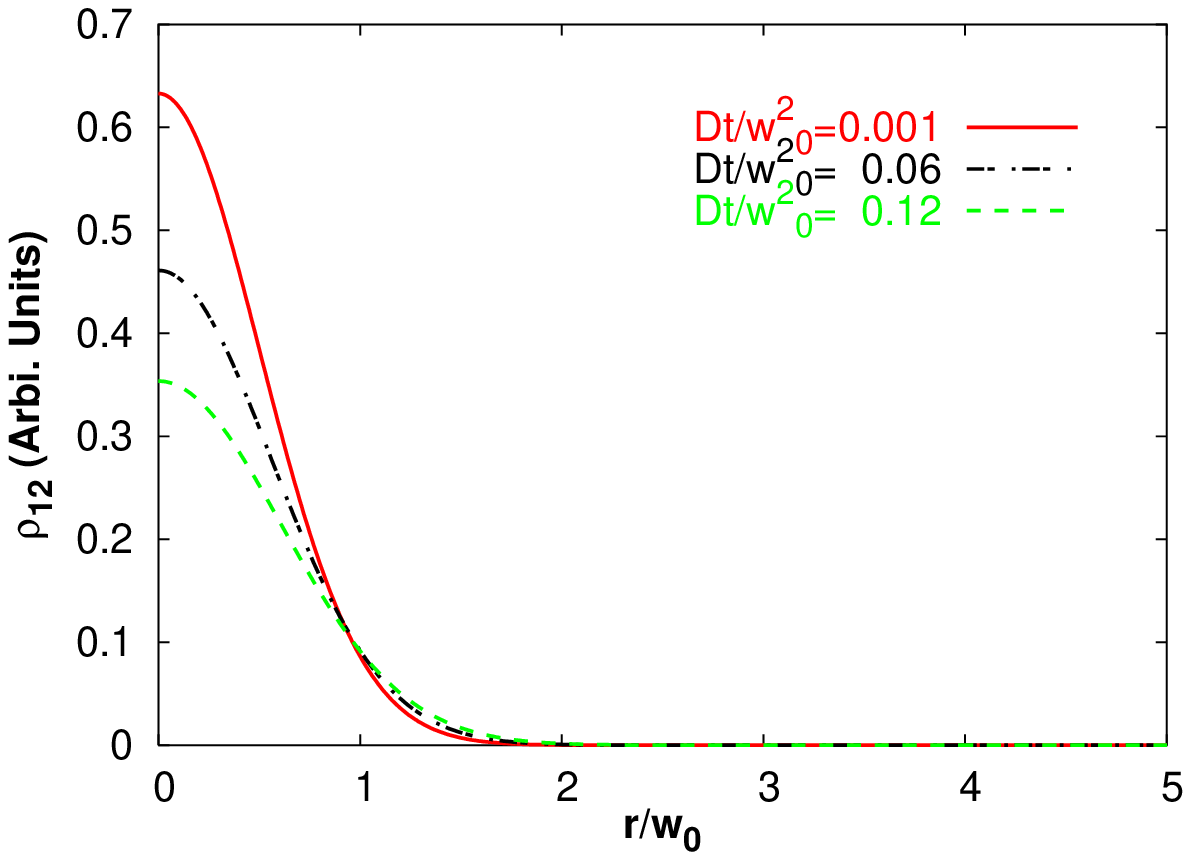}}
\centerline{\includegraphics[clip,width=.7\linewidth]{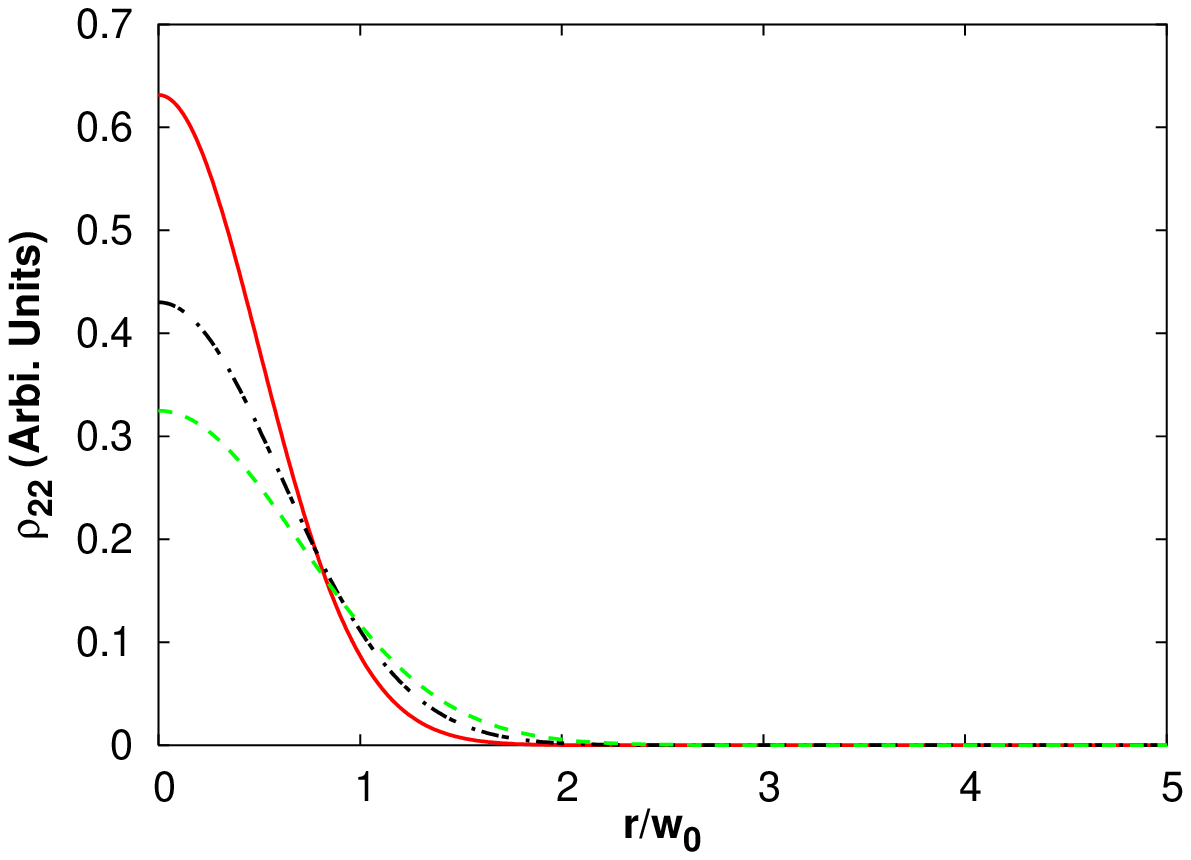}}
\centerline{\includegraphics[clip,width=.7\linewidth]{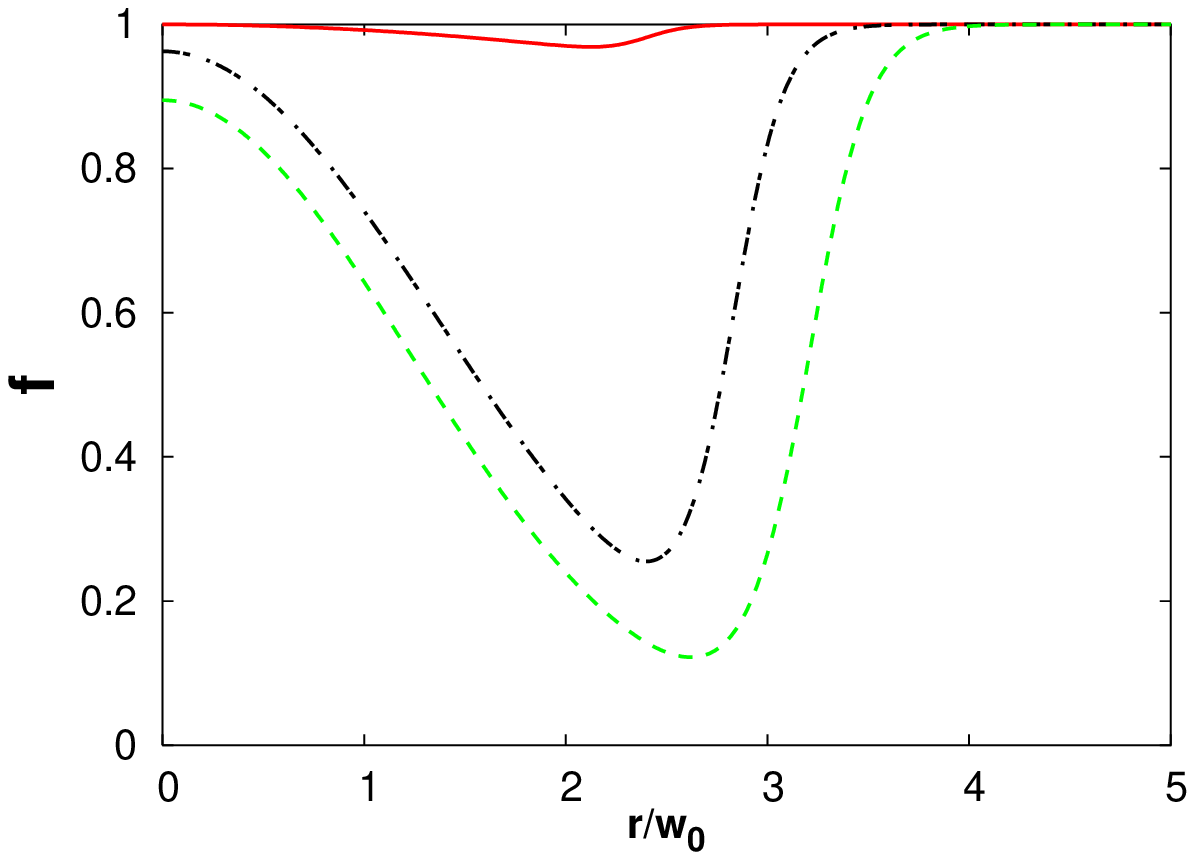}}
        \caption{\protect\label{Gaussian}  (Color online) Effects of
diffusion on a stored Gaussian mode. Plotted are coherence
$\rho_{12}$ (a), population $\rho_{22}$ (b) and coherence factor
$f$ (c) as functions of radius, for different diffusion times.
$w_0$, $D$ are the waist  of the Gaussian mode and the diffusion
coefficient, respectively.}
\end{figure}

{\em Comparison between the decay for vortex states and
vortex-free states} Using the propagator of the diffusion
equation, the expected atomic coherence of a LG mode after
diffusion of duration $t$ is~\cite{VortexDiffusion}
\begin{equation}
\label{eq:Coherence}
\rho_{12}\left(\vec{r},t\right)=\frac{(-g/\Omega)}{\sqrt{s(t)^{m+1}}}A_m(r,\sqrt{s(t)}w_0)e^{-im\theta}
\end{equation}
with $g$ the vacuum Rabi frequency for the probe field, $\Omega$
the Rabi frequency for the pump laser, $s(t)=(w^2_0+4Dt)/w^2_0$ an
evolution factor, $w_0$ the waist, and the radial profile of a LG
mode $A_m(r,w_0)=(1/w_0)\sqrt{2P/\pi
m!}(\sqrt{2}r/w_0)^m\exp(-r^2/w^2_0)$ where $P$ is the total
intensity. Note that Eq.~(\ref{eq:Coherence}) applies to the
Gaussian mode ($m=0$) as well. For example, the evolution of
stored coherence is shown in Fig.~\ref{vortex}a and
Fig.~\ref{Gaussian}a for $m=1$ and $m=0$, respectively. At large
$r$s, the coherence is homogeneously zero due to the exponential
factor in $A_m$. Note that Eq.~(\ref{eq:Coherence}) describes both
a spread of the coherence, indicated by $\sqrt{s(t)}$ inside
$A_m$, and a decay of the coherence compared a purely coherent
spread. The decay is given by $\sqrt{s(t)^{m+1}}$ in the
denominator. Then $F=1/s(t)^{m+1}$ gives the fidelity of the
stored coherence. $F$ shows that different angular momentum states
$m$s have different coherence decay factors. Although we will come
back to the decay of coherence later, we note here that the larger
OAM $m$, the larger the decay factor. The Gaussian mode, which has
no phase singularity, has smaller coherence decay factor than all
vortex states! This is because without vortex, the coherence is
always in phase for different locations and there is no
destructive interference to destroy the coherence. One can show
that the retrieval efficiency along forward direction is the same
as the fidelity $F$. Therefore, we come to a counter-intuitive
result --- a Gaussian state has a higher storage fidelity (or
retrieval efficiency) than the LG modes.

This may make the applications of optical vortex (with storage)
for QI questionable. Even though the excitation loss of stored
Gaussian or optical vortex state can be identified as a detected
error and the threshold for detected errors allowed by quantum
computation are realistic ~\cite{QClargeError},
\commentout{QCerror1,QCerror2} fast quantum computation with
reasonable resource overhead still requires relatively low error
rates. This is also true for quantum
communication~\cite{QComm1,LiangRepeater07}, where, although the
excitation loss can be fully controlled by its intrinsic
purification, a fast quantum repeater still needs a fairly high
retrieval efficiency.

The scaling factor $\sqrt{s(t)}$ in $A_m$ of
Eq.~(\ref{eq:Coherence}) means that the functional forms of
coherence are preserved for both the LG mode and the Gaussian
mode. This implies the functional form stability of stored
coherence against diffusion does {\em not} require topological
defects. Indeed, the disappearance of the dark center of the
center-blocked Gaussian mode after
diffusion~\cite{VortexDiffusion} demonstrates as the stability of
the Gaussian mode: diffusion tries to restore the nonzero
intensity at its center. Of course, the restoration can also be
understood by decomposing the center-blocked Gaussian mode to
LG$^0_p$ and noting that $p\neq0$ modes decay faster than $p=0$
mode and what is left after some time of diffusion is just the
Gaussian mode.

From the above discussion, it is clear that for the LG light
($m\neq0$), the coherence at the center $r=0$ stays zero during
diffusion. However, as we show now, population at $r=0$ does not
stay zero. To simplify the discussion, we assume, at time $t=0$,
the populations of atoms are $\rho_{11}(\vec{r},t=0)=1$ and
$\rho_{22}(\vec{r},t=0)=|\rho_{12}(\vec{r},t=0)|^2$, which is a
good assumption for strong pump and weak probe lasers as usually
used in light-storage experiments. At time $t$ after diffusion, we
have $\rho_{11}(\vec{r},t)=1$ and
\begin{equation}
\label{eq:Population}
\rho_{22}(r,t)=\frac{4e^{-\frac{2r^2}{8Dt+w^2_0}}P\left(32D^2t^2+r^2w^2_0+4Dtw^2_0\right)}{\pi\left(8Dt+w^2_0\right)^3}
\end{equation}
for $m=1$ vortex state and
\begin{equation}
\label{eq:PopulationG}
\rho_{22}(r,t)=\frac{2e^{-\frac{2r^2}{8Dt+w^2_0}}P}{\pi\left(8Dt+w^2_0\right)}
\end{equation}
for a Gaussian mode. The evolutions of populations are plotted in
Fig.~\ref{vortex}b and Fig.~\ref{Gaussian}b. While it seems that
coherence only diffuses outwards in Fig.~\ref{vortex}a,
Eq.~(\ref{eq:Population}) and Fig.~\ref{vortex}b clearly show that
diffusion goes in all directions as it should be. The outwards
moving coherence during diffusion is because the interference
cancels the inwards diffusing coherence. In contrast, the
population does not interfere with each other and thus the
diffusion towards the center is clearly seen. Indeed, the
population at the center quickly approaches a global maximum as
time increases (Fig.~\ref{vortex}b). We also note that integration
of $\rho_{22}$ over the whole space is conserved during the
diffusion.

{\em Diffusion induced decoherence} We have seen that diffusion
for coherence and population obeys different equations
(Eq.~(\ref{eq:Coherence}-\ref{eq:PopulationG})). This brings phase
decoherence for the stored coherence. To characterize the
decoherence, we define a coherence factor
$f=\frac{|\rho_{bc}|^2+\eta}{\rho_{bb}\rho_{cc}+\eta}$, where
$\eta$ is an infinitesimal number. $f$ is a linear function with
$f=1$ for a pure state and $f=0$ for a completely mixed state.
Thus, $f$ is a good parameter to describe the (local) coherent
property. As a specific example, we plot the coherence factor $f$
in Fig.~\ref{vortex}c and Fig.~\ref{Gaussian}c.
Figure~\ref{vortex}c shows that right after diffusion begins,
coherence factor $f(r=0)$ of the stored vortex drops from $1$ to
$0$ because at $r=0$, the population becomes nonzero when
diffusion starts while its coherence keeps zero. We note that such
sudden changes within a infinitesimal time are very uncommon in
physical processes. As diffusion time increases, $f(r)$ drops
faster for $r$s that do not have much initial inhomogeneous
coherence, and $f(r)$ approaches zero at very large time. This
latter result holds for a Gaussian mode as well (shown in
Fig.~\ref{Gaussian}c). We also note that at $r$s where the
diffusion of stored inhomogenous coherence and population has not
arrived at, the coherence factor $f$ stays at $1$ because all
population is in \ket1 and it is a pure state. But the weight
$\rho_{22}$, justified by its non-negative and conserved
integration, of these coherence factors for the retrieved light
approaches zero.

Here is how coherence factor $f$ may be obtained from the
experiments. When the reading pulse is applied, the incoherent
part will be retrieved as fluorescence in all directions.
Collecting both the fluorescence and the coherent emission then
allows extracting $f$. Of course, setting the detector at the
forward direction as generally used, e.g., in
~\cite{VortexDiffusion}, can only collect a finite ratio of
fluorescence, while the coherent part is collected by the
detector~\cite{DSP1}. What we want to emphasize is that
incoherence makes intensity at the center of a retrieved vortex
{\em nonzero}, prohibited by a coherent vortex state. As diffusion
time increases, nonzero intensity at the center builds up.
Therefore, generation of mixed state makes additional loss of
retrieved fidelity compared with vortex-free case. Diffusion of
the population makes visibility decrease and finally kill the
vortex. Of course, this part of reduction of fidelity can be
alleviated by using spatial filtering of the optical mode to
prevent spontaneously emitted photons from going into the
detectors. In this case, the retrieval probability is just the
fidelity $F=1/s(t)^{m+1}$.

The different collection efficiency of fluorescence photons and
coherent photons by a forward set detector helps explain why the
visibility of the center-blocked Gaussian mode disappears very
quickly, while the hole of vortex disappears very
slowly~\cite{VortexDiffusion}. The homogeneous phase of  the
center-blocked Gaussian mode makes nonzero coherence inside the
hole after diffusion and thus the disappearance of hole once the
coherence is read. This is in contrast with a stored vortex, whose
coherence at the center remains zero all the time. Its nonzero
intensity at the center comes only from incoherence of the center.
However, were most of fluorescence photons collected by the
detector, the dark center of a vortex would disappear quicker than
that of the Gaussian mode. It is the spatial filtering of the
optical mode that helps to overcome the fluorescence from the
center.

{\em Dependence of decay on order of phase singularity} We now
come back to the decay of coherence. We have noted that the
coherence of stored LG modes decays according to a power law
$F=1/s(t)^{m+1}$. The larger the order of phase singularity, given
by $m$, is, the faster the coherence decays. An additional example
of exponential decay rate $2D k^2$ due to diffusion of a plane
wave $e^{-i k x}$, which is faster than the power law decay, also
corroborates this idea, because a plane wave has infinity number
of phase singularities. Incidently, the limit of $F$ for large $m$
does not go to exponential decay of a plane wave is because the
amplitude of coherence in LG modes is not a constant, different
from the case of a plane wave. We also note that the larger $k$ of
a plane wave, the larger decay rate, which is because larger $k$
means the pattern have higher spatial frequency and the diffusion
becomes easier to cancel the coherence. Such diffusion of a plane
wave happens if the pump and probe lasers have different wave
vectors.

A few more remarks on the different decay behaviors are in order.
First, as a direct application of our discussion, the less phase
gradient, the better for stored coherence against diffusion. For
example, diffusion of stored general LG$^m_p$ with both winding
number and number of radial nodes being non-zero~\cite{LGintro}
induces faster decay for $p>0$ than $p=0$. Furthermore, we note
that although the number of nodes in the coherence does not change
with diffusion time, the positions for the off-center radial nodes
change, which is different from the center one. The non-moving
position of the center node comes from geometric symmetry of the
vortex. Since the decay rate depends on both $|m|$ and $p$ for the
mode LG$_{p}^{m}$, if OAM states are to be used as bases for
quantum information, a preferred basis to reduce loss of
entanglement for quantum information is actually two modes with
same $p$ but opposite $m$. Second, since diffusion induced decay
of the total intensity is not exponential for Gaussian and LG
modes, the exponential decay at a rate $20,000 s^{-1}$ of
Ref.~\cite{VortexDiffusion} is not determined by diffusion.

{\em The classical diffusion vs the quantum diffusion} %(MD06/05/07):
So far, we have only discussed decoherence induced by the {\em
classical} diffusion associated with the inhomogeneous
distribution of coherence. We note that decoherence can also
happen in a homogeneous system as a result of the quantum
diffusion. The quantum diffusion in light storage system happens
when pump and probe lasers couple different momentum states, which
introduces decoherence~\cite{TunRamanSuper}. But this decoherence
is reversible, say by photon echo techniques, in contrast with the
classical diffusion. This is because the quantum diffusion is
described by the complex Schrodinger equation with $i$ in it,
while classical diffusion does not. Finally, we note that when the
same momentum states are coupled by choosing pump and probe lasers
to have the same wave vectors, the quantum diffusion
disappears~\cite{TunRamanSuper}. Incidently, inhomogenous magnetic
fields also induce quantum diffusion.

{\em Discussion} Our results indicate that diffusion actually
introduces more decoherence in a stored vortex mode than a stored
Gaussian mode, which may have important implication for quantum
information. Of course we, however, do not rule out that in other
processes, such as a quantum gate operation, vortex states are
possibly much better than the topological-free state. Nor did we
discuss diffusion-free systems such as BEC~\cite{VortexStoreBEC}
and bound excitons in semiconductors~\cite{KaiMei05,TunEITBX}.

{\em Conclusion} We found that during diffusion, the coherence of
stored vortex states decays faster than that of Gaussian states.
This is surprising because vortex states are associated with
topological properties, and are presumably considered as more
stable than topological-defect-free states. The underlying reason
is that diffusion is a non-local process. More generally, the less
phase gradient of a stored coherence, the better for it against
diffusion. Furthermore, calculation of coherence factor showed
that the center of stored vortex becomes completely incoherent
once diffusion begins, and when reading laser is applied, the
optical intensity at the center of the vortex builds up. It's
implication for quantum information was discussed. Finally, we
compared the classical diffusion and the quantum diffusion.

This work is supported by NSF and Research Corporation. T. Wang
and L. Zhao acknowledge Yanhong Xiao for bringing
Ref.~\cite{VortexDiffusion} to their attention, and J. Javanainen
for his critical reading of the manuscript.

\bibliography{Diffusion}

\end{document}